# The impact of the following vehicles' behaviors on the car-following behaviors of the ego-vehicle

Yang Liu, Jiahao Zhang, Yuxuan Ouyang, Huan Yu, and Dengbo He, *Senior Member, IEEE*

*Abstract*— Among all types of crashes, rear-end crashes dominate, which are closely related to the car-following (CF) behaviors. Traditional CF behavior models focused on the influence of the vehicle in front, but usually ignored the peer pressure from the surrounding road users, including the following vehicle (FV). Based on an open dataset, the highD dataset, we investigated whether the FV's states can affect the CF behavior of the ego-vehicle in CF events. Two types of CF events were extracted from highD database, including the tailgated events, where the time headway between the FV and the ego-vehicle (i.e., time gap) was ≤ 1 second, and the gapped events, where the time gap was ≥ 3 seconds. The dynamic time warping was to extract CF pairs with similar speed profiles of the leading vehicle (LV). Statistical analyses were conducted to compare the CF-performance metrics in tailgated and gapped events. Then, the inverse reinforcement learning was used to recover the reward function of the ego-vehicle drivers in different CF events. The results showed that the ego-driver would adjust their CF behavior in response to the pressure from a tailgating FV, by maintaining a closer distance to the LV, but at the same time, driving more cautiously. Further, drivers were still able to adjust their CF strategies based on the speed of traffic flow and the distance to the LV, even when being tailgated. These findings provide insights regarding more accurate modelling of traffic flow by considering the peer pressure from surrounding road users.

## I. INTRODUCTION

THE rear-end collisions, accounting for a large proportion of all crashes on the road, have posed a major threat to traffic efficiency and safety [1-3]. The rear-end collisions are closely related to drivers' car-following (CF) behavior. Thus, various driving behavior models have been proposed to characterize drivers' decision-making processes and vehicle dynamic responses in CF events, including model-driven approaches based on driving behavior theories, such as the Intelligent Driver Model (IDM) [4] and Gazis-Herman-Rothery (GHR) model [5], and data-driven approaches that rely on large-scale traffic data, including deep learning and reinforcement learning methods [6-9]. However, to the best of our knowledge, existing driver behavior models mostly assume drivers only look forward, i.e., making CF decisions based on the states (e.g., speed, acceleration and position) of the lead vehicles (LVs), in order to maintain traffic safety and flow efficiency [10, 11].

At the same time, it is well-known that drivers would actively seek information from the traffic environment to act proactively – not just the information directly ahead of the ego-vehicle but also the information from other directions [12]. For example, drivers may consider the intention of pedestrians beside the road and brake in advance before the LV brakes [13]. In the CF event, drivers may also consider more than the LV information. For example, aggressive drivers have widely adopted tailgating to nudge the leading drivers to yield the lane or drive faster. When being tailgated, drivers may notice the states of the approaching vehicle (e.g., relative speed and distance) from the rear-view mirrors and adjust their speed control strategies. However, to the best of our knowledge, the effect of the following vehicles (FVs) on the CF behavior of ego-vehicles has not been quantified.

Thus, to quantitatively analyze the impact of the FV's state on the ego-vehicle driver's behavior, we extracted CF segments from the highD naturalistic driving dataset [14]. The FV states were classified into two categories: tailgated and gapped, based on the headway between the FV and the ego-vehicle. Specifically, if the time headway was 1 second or less, the scenario was classified as tailgated, indicating a high-pressure driving condition [15]. Conversely, if the headway was 3 seconds or more, the scenario was categorized as gapped, representing a low-pressure condition with increased maneuvering freedom for the ego-vehicle driver [16]. We then compared the differences in driving behavior under the two conditions, with a focus on speed fluctuation and driving safety. Next, the Adversarial Inverse Reinforcement Learning (AIRL) [17] algorithm was employed for the tailgated and gapped scenarios. Based on the reward function generated by the AIRL model, we further quantified the influence of the FV on the ego-

Submitted on July 1, 2025. This work was supported by the Scientific Research Projects for the Higher-educational Institutions of Guangzhou (No. 2024312135), and in part by the Guangzhou Municipal Science and Technology Project (No. 2023A03J0011), and Guangdong Provincial Key Lab of Integrated Communication, Sensing and Computation for Ubiquitous Internet of Things (No.2023B1212010007). (Corresponding author: Dengbo He).

Yang Liu (e-mail: yliu068@connect.hkust-gz.edu.cn) and Jiahao Zhang (e-mail: jzhang012@connect.hkust-gz.edu.cn) are with Intelligent Transportation Thrust, The Hong Kong University of Science and Technology (Guangzhou), Guangzhou, China.

Yuxuan Ouyang is with College of Educational Science, The Hong Kong University of Science and Technology (Guangzhou), Guangzhou, China (e-mail: youyang354@connect.hkust-gz.edu.cn).

D. Yu is with Robotics and Autonomous Systems Thrust and Intelligent Transportation Thrust, The Hong Kong University of Science and Technology (Guangzhou), Guangzhou, China (e-mail: huanyu@hkust-gz.edu.cn).

D. He is with the Intelligent Transportation Thrust and Robotics and Autonomous Systems Thrust, The Hong Kong University of Science and Technology (Guangzhou), Guangzhou, China and the HKUST Shenzhen-Hong Kong Collaborative Innovation Research Institute, Guangdong, China (e-mail: dengbohe@hkust-gz.edu.cn).

vehicle driver's CF decisions. These findings can provide valuable insights into how drivers adapt their behavior in response to different FV behaviors, and provide implications for traffic safety, driver assistance system design, and CF behavior modeling in real-world traffic environments.

## II. BACKGROUND

### A. CF Controllers with FV State Considered

The states of surrounding vehicles have been considered to improve the performance of CF controllers. For example, Wang, et al. [1] used the Soft Actor-Critic (SAC) algorithm-based CF controller to improve the safety of human drivers, with the states of the LVs and FVs considered, including the speed of the ego-vehicle, the relative speed and the relative distance between the ego-vehicle and LV, and between the ego-vehicle and the FV. Similarly, Li, et al. [18] proposed a method to quantify the influence of surrounding vehicles on the CF behaviors of the ego-vehicle and found that with more surrounding vehicle information considered (e.g., headway between two adjacent vehicles, location of the n-th vehicle, and speed difference between two adjacent vehicles), the risk of traffic collisions can be reduced. Quantitatively, Peng, et al. [19] also proposed an improved CF model based on the full velocity difference (FVD) controller that considered the states of both the LV and the FV vehicles in CF events, including speed, position, and acceleration of LV and FV. They found that when calibrated on a real-world dataset, the new model can achieve stable following in a shorter time and improve the efficiency of traffic flow compared to traditional FVD model. On top of FVD controller, Zong, et al. [20] proposed a CF controller, the Full Velocity and Acceleration Difference (FVAD), which took into account the states of multiple LVs and one FV through molecular dynamics theory. Optimized by field experiment data, the FVAD was found to be more stable than the Adaptive Cruise Control (ACC) controller.

### B. Endeavors to Reduce Rear-End Collisions

As rear-end collision accounts for a large proportion of traffic accidents in both vehicles with and without driving automation [21-23], and thus endeavors have been made to improve CF safety. On the one hand, various human-machine interfaces (HMIs) have been proposed to enhance drivers' perception of the traffic ahead of the ego-vehicle in order to reduce the chance of ego-vehicle rear-ending the LVs. For example, Yan, et al. [24] proposed an HMI showing the beyond-visual-range information of the leading traffic flow (i.e., showing the motion states of the LV of the vehicle directly ahead) and found that such HMI can facilitate earlier responses to the deceleration of the traffic flow ahead. On the other hand, researchers also tried to enhance drivers' perception of rear events in order to reduce the chance of being rear-ended in traffic flow. For example, through a driving simulator experiment, Roessing, et al. [25] found that providing the distance, speed, and risk of FVs in the rear-view mirror can enable ego-vehicle drivers to better sense the risk of FVs. Flannagan [26] compared the optical and video-stream-based rear-view mirrors and found that the average perception distance of following hazards was 27.2 m with optical mirrors and 35.4 m with video-stream-based mirrors, indicating the safety benefits of video-stream-based rear-view mirrors.

### C. Research Gap and This Study

Previous studies have shown that considering the relative distance or speed between the FV and the ego-vehicle can enable better vehicle control strategies of the ego-vehicle. However, most of the research focused on the optimization of the CF controller design. Although they have highlighted the importance of considering FV's states in CF events, they could not reveal the actual behavior of human drivers in CF events. Given that previous research also highlighted the importance of perceiving the following traffic for drivers to better avoid being rear-ended, it would be necessary to further quantify the role of following traffic information on drivers' CF behaviors in order to better model drivers' CF behaviors, which is the objective of our study.

## III. DATA PREPARATION

### A. The highD Open Dataset

The CF events were extracted from the highD dataset [14], which is a 147-hour natural traffic flow dataset obtained by RWTH Aachen University, collected near the Cologne motorway in Germany in 2017-2018 using drones. The dataset includes 110,500 vehicles and covers a range of 420 m, with a vehicle trajectory tracking accuracy of 10 cm.

One of the key advantages of the highD dataset is its high precision in trajectory tracking, which enables reliable identification of different FV states. Additionally, its extensive coverage across various traffic densities and road conditions makes it possible to analyze CF behavior under diverse scenarios. The use of aerial drone footage eliminates occlusion issues commonly found in ground-based sensor data, providing an unobstructed view of vehicle interactions. This enables accurate measuring of time headway and relative speed in CF events, which serves as the foundation for the analysis of drivers' CF behaviors.

### B. Extraction of Car-Following Segments

Following previous research, we set the following criteria for the extraction of CF segments:

- The LV and the ego-vehicle need to be in the same road throughout the data extraction period to ensure continuous CF segment [27-30].
- The LV is within 100 m of the ego-vehicle to ensure the behaviors of the ego-vehicle can be affected by the LV [28].
- The speed of both LV and the ego-vehicle should be ≥ 10km/h to exclude traffic jams, given that we are more interested in drivers' CF behaviors in free-flowing traffic and drivers' behaviors might be different in stop-and-go traffic [31].
- The duration of CF segment should be at ≥ 10s to ensure the reliability of the CF data [27, 29, 30].

Specifically, as shown in Figure 1, we extracted two types of CF events, i.e., the tailgated and gapped CF event. As mentioned previously, according to the widely accepted three-second following rule, an event with a headway exceeding 3 seconds between the FV and the ego-vehicle was classified as a gapped CF events [32]. Further, as drivers were found to prefer a minimum headway of approximately 1 second during rush hour traffic [15], and on German highways, fines are imposed when the time headway falls below 0.9 seconds [33], we defined the CF events with a headway of 1 second or less as the tailgated CF events, where the FV tailgated the ego-vehicle. Finally, 1,024 tailgated CF events and 465 gapped CF events are gotten in this study.

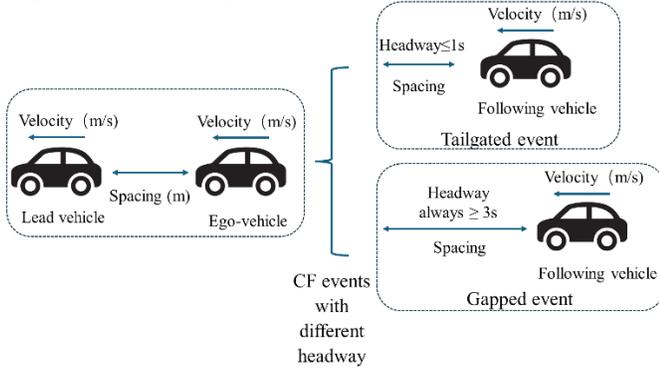

**Fig. 1.** Two types of CF Events.

## IV. DATA ANALYSIS

To ensure a fair comparison of CF behaviors CF events, we first used a Dynamic Time Warping (DTW) method to extract driving segments with similar LV speed profiles. Then, to illustrate the impact of tailgaters on the CF behavior of the ego-vehicle, we compared the metrics related to speed fluctuation and driving safety between the tailgated and gapped CF events, based on several commonly used metrics. Then, to quantify the CF decisions in different scenarios, we used Adversarial Inverse Reinforcement Learning (AIRL) to learn the dynamic behavior of ego-vehicles, enabling a quantitative representation of driver preferences across different situations.

### A. Dynamic Time Warping

DTW [34] is an algorithm commonly used to compare the similarity between time series. Specifically, the DTW can non-linearly align two time series to find the optimal match. Then, it uses Dynamic Programming (DP) [35] techniques to calculate a "distance measure," which quantifies the similarity between the two sequences. In this study, we matched the speed profiles of the LV in the CF segments using the DTW, which enables the analysis of differences in ego-vehicle's CF behavior under similar traffic flow conditions.

The specific implementation of DTW is as follows: Given two time series sequences $X = (x_1, x_2, \ldots, x_n)$ and $Y = (y_1, y_2, \ldots, y_m)$, where $m$ and $n$ represent the lengths of the two sequences, a matrix $D$ of size $m \times n$ is constructed. The element $D(i, j)$ represents the minimum cumulative distance between the subsequence $x_1$ to $x_i$ and $y_1$ to $y_j$. The recursive relationship for the matrix is defined as follows:

$$D(i, j) = d(x_i, y_j) + min[D(i - 1, j), D(i, j - 1), D(i - 1, j - 1)] \quad (1)$$

Where $d(x_i, y_j)$ represents the Euclidean distance between $x_i$ and $y_j$ in two sequences. The lower-right element $D(n, m)$ of the matrix D is the DTW distance between the two sequences

- the smaller the distance, the more similarity between the sequences.

### B. Driving Behavior Metrics

#### a) Speed Fluctuation

The speed variation of the ego-vehicle in a CF segment can be manifested as the frequency of vehicle acceleration and deceleration [36, 37], which can impact the stability of traffic flow [38, 39]. The metrics of speed variation are as follows:

**The standard deviation of the vehicle speed in a CF segment (Std)**: a statistic that measures how discrete or dispersed the data points are. The equation of $Std$ is shown below:

$$Std = \sqrt{\frac{\sum_{i=1}^{n}(v_i - \bar{v})^2}{n-1}} \quad (2)$$

Where, $v_i$ is the sample point of speed in a dataset; $\bar{v}$ is the mean value of speed samples; and $n$ is the sample size in the dataset

**Mean Absolute Deviation (Dmean)**: a statistic that measures the degree to which a data point is off-center (usually the mean or median). Unlike standard deviation, $D_{mean}$ does not square outliers, so it reflects the actual distribution and fluctuations of the data better than standard deviation. The equation of $D_{mean}$ is shown below.

$$D_{mean} = \frac{\sum_{i=1}^{n}|v_i - \bar{v}|}{n} \quad (3)$$

The meaning of $v_i$, $\bar{v}$, and $n$ is the same in the Std equation.

**The coefficient of variation (Cv)**: $C_v$ measures the relative speed variability in a CF segment. It is the ratio of the standard deviation to the mean. The equation of $C_v$ is shown below.

$$C_v = \frac{Std}{|\bar{v}|} \times 100\% \quad (4)$$

Where, $\bar{v}$ is the mean speed in a CF segment.

**The time-varying stochastic volatility (Vf)**: $V_f$ can describe the conditional heteroscedasticity of the speed. It measures the fluctuation of the speed sample by computing the changes in the proportion of observations. The equation of $V_f$ is shown below.

$$V_f = \sqrt{\frac{\sum_{i=1}^{n}(r_i - \bar{r})^2}{n-1}} \quad (5)$$

Where, $r_i = ln\left(\frac{v_i}{v_{i-1}}\right) \times 100\%$; $v_i$ and $v_{i-1}$ are the observed speed sample $i$ and $i - 1$, respectively; $\bar{r}$ is the mean value of $r_i$; and $n$ is the sample size.

#### b) Driving safety

**The time headway (HWY)**: HWY is the time interval between the heads of two adjacent vehicles, usually expressed in seconds [15] and calculated as below:

$$h_w(t) = \frac{X_L(t) - X_E(t)}{V_E(t)} \quad (6)$$

Where $X_L(t)$ and $X_E(t)$ are the position of LV and SV at time t; and $V_E(t)$ is the speed of ego-vehicle at time t.

**Deceleration Needed to Avoid Crash (DRAC)**: DRAC is a metric that quantifies the minimum deceleration required for the vehicle to avoid a collision with the vehicle in front, assuming the vehicle brakes at a constant deceleration until it comes to a complete stop. This index is calculated based on kinematic equations and assumes uniform deceleration of the vehicle during braking [40]. To avoid a collision, the ego-vehicle must decelerate fast enough to match the speed of the

LV in front and before crashing. The calculation of DRAC is as follows:

$$\begin{cases} DRAC(t) = \dfrac{(V_E(t) - V_L(t))^2}{2(X_L(t) - X_E(t) - L)}, & if \ V_E(t) > V_L(t) \\ \qquad\qquad 0, & otherwise \end{cases} \quad (7)$$

Where, $X_L(t)$ and $X_E(t)$ are the positions of LV and ego-vehicle at time $t$; $V_L(t)$ and $V_E(t)$ are the velocity of LV and ego-vehicle at time $t$; and $L$ is the length of the LV. The larger the DRAC value, the risker the scenario.

### C. Adversarial Inverse Reinforcement Learning (AIRL) to Recover Rewards in CF Events

#### a) Modeling CF as a Markov Decision Process (MDP)

The CF process can be recognized as an MDP [41]. In MDP, the object being controlled is called an agent, the surroundings with which the agent interacts are the environment, and the interaction between the agent and the environment is the action. An MDP can be described by five elements: i) the state, $S$, a description of the environment, which can be discreet or continuous. ii) the action, $A$, a collection of all possible movements that an agent can perform at each state and defines how the agent affects the environment. iii) the transition probability, $P(s'|s, a)$, which is the probability of moving to the next state $s'$, given the current state $s$ and the action $a$. Transition probability reflects the dynamic behavior of the environment; iv) the reward, $R$, an assessment of the environment. The reward can be positive (good state) or negative (bad state), and it guides the learning direction of the agent; v) the discount factor, $\gamma$, a constant between 0 and 1, indicating the importance of future rewards. The discount factor makes future rewards less important than present rewards.

The interaction between the agent and the environment is the choice of actions in a certain state, and this choice is called policy ($\pi$). Rewards affect the agent's policy. However, if only the current and next state rewards are considered, the policy will fall into local optimality and fail to reach global optimality. So, when training MDP model, one should take into account the present state and all possible future states. Reinforcement learning (RL) [42] is a process to learn an optimal strategy under the MDP framework so that the agent can maximize its accumulated discount reward through interaction with the environment. Specifically, the goal is to find a $\pi$ that selects the optimal action in each state, thereby maximizing long-term returns.

In our study, the agent of RL model is the ego-vehicle and the agent action is acceleration or deceleration decision. The states can be defined as a vector $[\Delta y_{le}, v_e, v_l, \Delta v_{le}]$, composed of relative distance $\Delta y_{le}$ between LV and ego-vehicle, ego-vehicle velocity $v_e$, LV velocity $v_l$, and relative speed $\Delta v_{le}$ between LV and ego-vehicle. According to Newton's kinematic laws, the physical relationships between these state parameters follow the following equations:

$$v_e(t+1) = v_e(t) + a(t) \times \Delta T \qquad (8)$$

$$y_e(t+1) = y_e(t) + \frac{1}{2}[v_e(t) + v_e(t+1)] \times \Delta T \qquad (9)$$

$$\Delta v_{le}(t+1) = v_l(t+1) - v_e(t+1) \qquad (10)$$

$$\Delta y_{le}(t+1) = y_l(t+1) - y_e(t+1) \qquad (11)$$

#### b) Adversarial Inverse Reinforcement Learning

AIRL is a method that combines generative adversarial network (GAN) [43] and inverse reinforcement learning (IRL) [44] to infer the reward function of the environment from the behavior of agents. As shown in Figure 2, the structure of AIRL is same as Generative Adversarial Imitation Learning (GAIL) [45], but the discriminator design is different. Specifically, in AIRL, a pseudo-trace was generated and used in the discriminator, $D_{\theta,\phi}(s, a, s')$, with the value ranging from 0 to 1 [17]. When the value of the pseudo-trace approaches 1, it is considered as a true trajectory in the discriminator; whereas when value of the pseudo-trace approaches 0, it is considered a false trace in the discriminator.

During policy optimization, Proximal Policy Optimization (PPO) [46] was adopted as the RL algorithm. We used the reward $\hat{r}(s, a)$ computed from the discriminator $D_{\theta,\phi}(s, a, s')$ to train the agent, which in turn generated pseudo-traces. The discriminator $D_{\theta,\phi}(s, a, s')$ was composed of two neural networks, as shown in Figure 2, denoted as $g_\theta$ and $h_\phi$, with network parameters $\theta$ and $\phi$, respectively. In the loss function $L(\psi)$, $\rho E$ denotes the occupancy measure of the expert policy, and $\rho\pi$ denotes the occupancy measure of all state-action pairs $(s, a)$ under the learned policy. The goal of $L(\psi)$ was to minimize the discrepancy between $\rho\pi$ and $\rho E$, thereby encouraging the policy to imitate expert behavior through interaction with the environment. To achieve this, the agent must interact with the environment to obtain the next state and make subsequent decisions. Under this training scheme, the optimal solution of the discriminator $D_{\theta,\phi}(s, a, s')$ corresponds to the solution of Maximum Entropy Inverse Reinforcement Learning (MaxEnt IRL) [47]. Accordingly, the output of the $g_\theta$ network represents the true reward function $R(s)$.

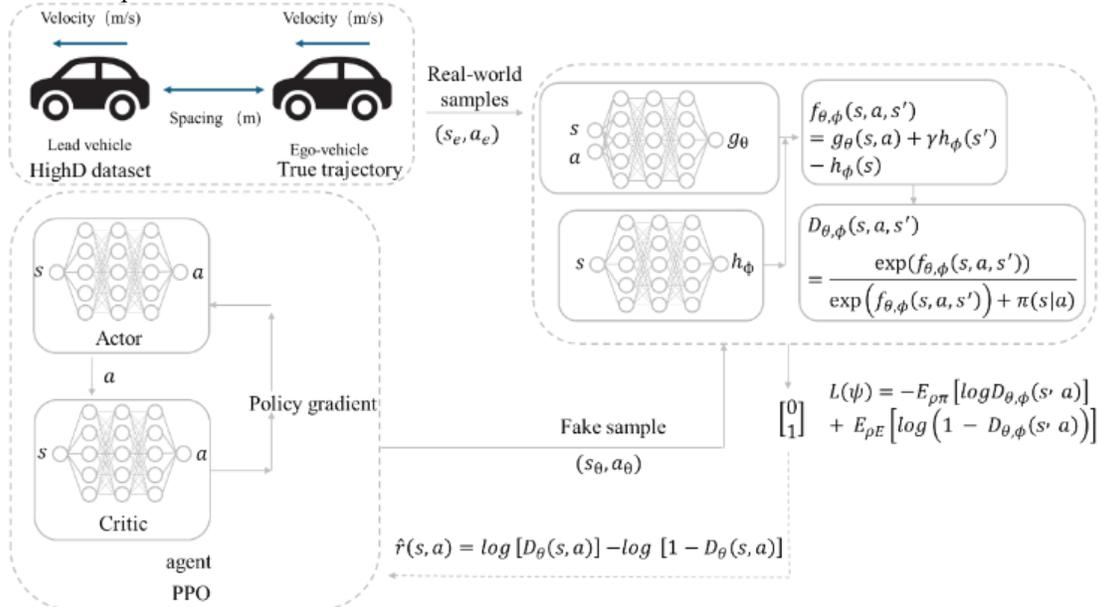

**Fig. 2.** The structure of AIRL.

The AIRL algorithm is a type of imitation learning method. During training, the stability of the learning process is evaluated by measuring the discrepancy between the generated trajectories and the real trajectories. This discrepancy is referred to as the training loss, and the equation of training loss is as follow:

$$loss = log\left(\frac{V_{et} - V_{em}}{V_{et}}\right) - 1000 \times c \quad (12)$$

Where $V_{et}$ was the true trajectory's driving speed of ego-vehicle in the CF segment, $V_{em}$ is the velocity of the ego-vehicle generated by the model, and $c$ is 1 if a collision occurred (otherwise 0). After the model was trained, the $g_\theta$ network was used to generate the reward function of the agent. Practically, this network took state variables as input, including the spacing between the LV and the ego-vehicle, the relative speed between the LV and the ego vehicle, and the speed of the ego-vehicle and then output a corresponding reward value.

## V. RESULTS

### A. Segments with Similar LV Speed Profiles

Based on DTW pairing, combined with manual screening and validation, we extracted 81 pairs of CF segments from the highD dataset, where the LV speed profiles were similar but the ego-vehicle was either tailgated or gapped by the FV. We visualize two examples of the paired results below:

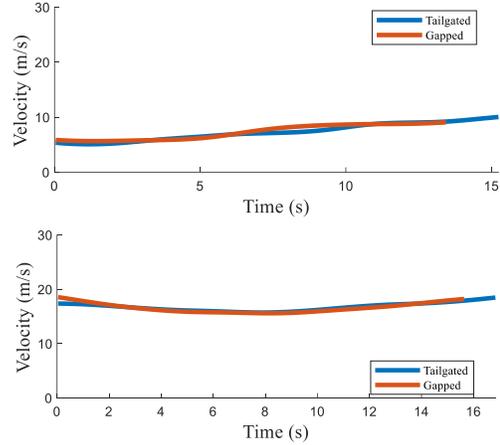

**Fig. 3.** LV speed profiles in paired CF events.

### B. Comparisons of CF Kinematic Metrics

First, we visualize the CF behavior of ego-vehicle when it was being tailgated or gapped by the FV. As shown in Figure 4, the kinematic indicators included bumper-to-bumper spacing between the ego-vehicle and the LV, ego-vehicle acceleration, and relative velocity between the LV and ego-vehicles. These variables are the inputs of established CF models, such as IDM and GHR. To facilitate better interpretation of CF behavior under varying conditions, we adopted two-dimensional kernel density estimation plots to visualize the distribution of kinematic metrics against the speed of the LV.

### C. Statistical Analysis of the CF Aggressive and Safety Metrics

In this section, a statistical analysis was conducted to further compare the CF-related metrics. Specifically, we calculated the mean, maximum, minimum, and standard deviation (SD) of each metrics for the paired CF segments and conducted paired t-tests between the paired samples.

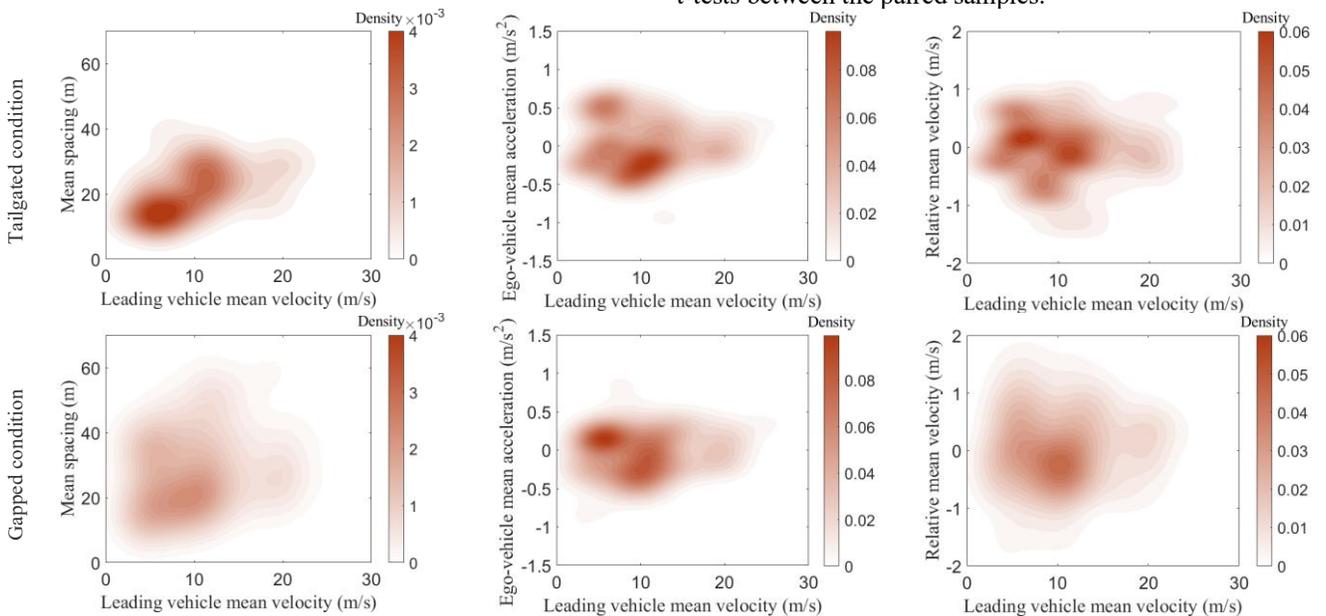

**Fig. 4.** Driver behavior density heatmap for paired car-following segments. The color intensity represents sample density, with darker colors indicating a higher density.

TABLE I.
COMPARISON OF CF-RELATED METRICS BETWEEN TAILGATED AND GAPPED CF EVENTS.

| Metrics | Tailgated CF Events | | | | Gapped CF Events | | | | Δ (%) | p-value |
|---|---|---|---|---|---|---|---|---|---|---|
| | Max | Min | Mean | SD | Max | Min | Mean | SD | | |
| | *Speed Fluctuation* | | | | | | | | | |
| $V_{sd}$ (m/s) | 3.83 | 0.13 | 1.16 | 0.68 | 2.97 | 0.35 | 1.17 | 0.55 | -0.86 | .8 |
| $D_{mean}$(m/s) | 3.21 | 0.12 | 0.98 | 0.58 | 2.77 | 0.28 | 0.99 | 0.50 | -1.02 | .6 |
| $C_v$ | 37.38 | 0.64 | 13.62 | 8.74 | 37.64 | 4.35 | 15.72 | 7.23 | -12.11 | .8 |
| $V_f$(m/s) | 0.37 | 0.02 | 0.86 | 0.09 | 2.08 | 0.58 | 1.04 | 0.31 | -20.93 | .07 |
| | *Driving Safety* | | | | | | | | | |
| Mean THW (s) | 5.11 | 0.77 | 2.48 | 1.06 | 27.92 | 0.83 | 3.67 | 2.82 | -47.98 | <.001* |
| Mean DRAC(s) | 0.05 | $1\times10^{-4}$ | 0.015 | 0.03 | 0.05 | $3\times10^{-4}$ | 0.02 | 0.03 | -33.33 | .03* |

*Note: In the table, * marks significant results (p < .05); Δ is the percentage difference between two types of CF events, i.e., Δ = (Tailgated Events − Gapped Events)/Gapped Events; THW is the time headway, and SD stands for standard deviation.*

As shown in Table I, first, we found that the speed fluctuation of the ego-vehicle was similar regardless of the state of the FV, indicating that tailgaters had limited influence on the speed variation of the ego-vehicle. In other words, the ego-vehicle still in general followed the speed of the traffic flow, regardless of the pressure from the FV. However, we observed significant differences between the tailgated and gapped CF scenarios in terms of the safety indicators (Mean THW and Mean DRAC).

Specifically, a smaller Mean THW in tailgated events indicates that the driver kept a smaller safety buffer when they felt the pressure from behind. The smaller Mean DRAC in tailgated events, however, indicates that, though keeping a smaller safety buffer, drivers were able to adjust their driving strategies to become more cautious when handling a smaller following distance.

### D. Results of AIRL

Finally, to better quantify the CF behaviors when being tailgated or not, we present the results of the AIRL. As shown in Figure 5, regardless of FV states, more than 1200 epochs were needed to reach stability of the AIRL model. To evaluate the model performance, we compared the real trajectories of the ego-vehicle to the trajectories generated by the model in both tailgated and gapped events. Specifically, the comparisons between the real spacing and the model-generated spacing between the LV and ego-vehicle, and between the real speed and the model-generated speed of the ego-vehicle demonstrated that the AIRL model can replicate the CF behavior of the ego-vehicle, as visualized in Figure 6.

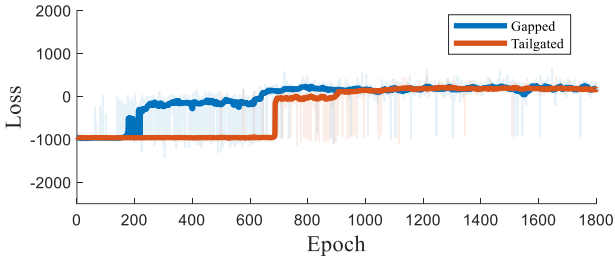

Fig. 5. Loss of AIRL training process.

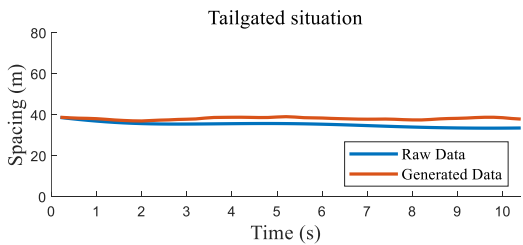

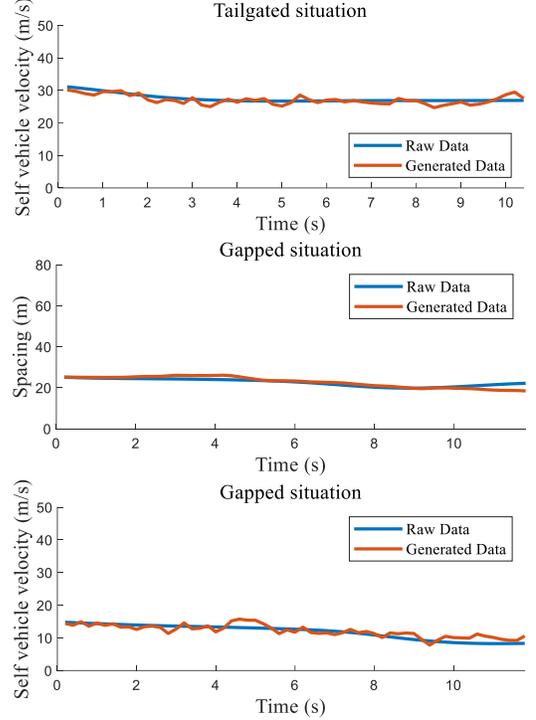

Fig. 6. A subset of the trajectories to illustrate the comparisons between AIRL-generated and real-world trajectories.

To analyze the behavioral differences of the ego-vehicle under different FV conditions, we computed the rewards for all possible states and compared the reward distributions across different scenarios. For better visualization, we discretized each state variable into ten evenly spaced ranges and generated corresponding reward heatmaps. In these heatmaps, a more yellow color (and thus larger reward value) indicates a higher reward value, suggesting that the driver had a stronger preference to stay at that particular state; while a more blue color (and thus smaller reward value) represents a lower reward, indicating driver's aversion to that state. The detailed results are presented in Figure 7.

In general, it was found that in the tailgated CF event, positive rewards predominantly occurred when the relative speed was negative, indicating that, in most cases, the ego-vehicle tended to move closer to the LV when the FV is in a tailgated state. Conversely, when the FV was in a gapped condition, the ranges of positive rewards for the ego-vehicle were larger on all dimensions, indicating that a larger following distance allows the ego-vehicle to adopt a more flexible and varied driving style.

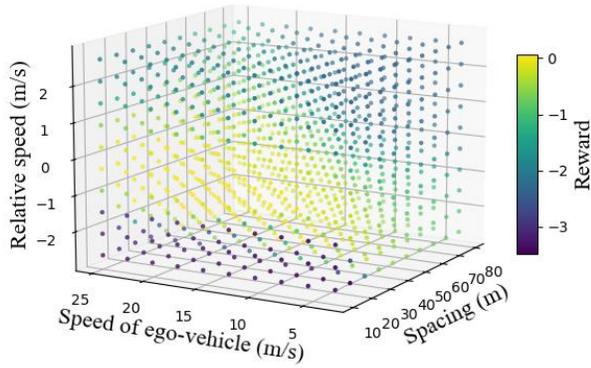

(a) Tailgated events

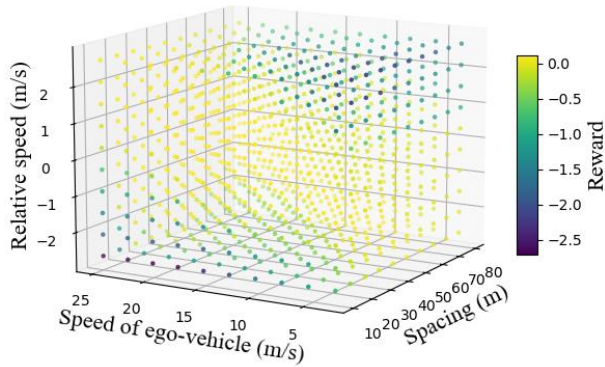

(b) Gapped events

**Fig. 7.** Reward for ego-vehicle CF decisions in different FV conditions.

To better demonstrate the impact of different FV states on the CF behavior of the ego-vehicle, we selected several representative LV speeds and compared the reward function of the ego-vehicle under different FV conditions based on the speed distribution of the paired CF segments. As shown in Figure 8, we can observe 4 distinct distribution centers when matching the speed of the ego-vehicle in tailgated and gapped events, i.e., 4.3 m/s, 7.4 m/s, 11 m/s, and 20 m/s. Thus, as shown in Figure 9, we visualized ego-vehicle driver's reward functions at these four speeds in tailgated CF events and gapped CF events.

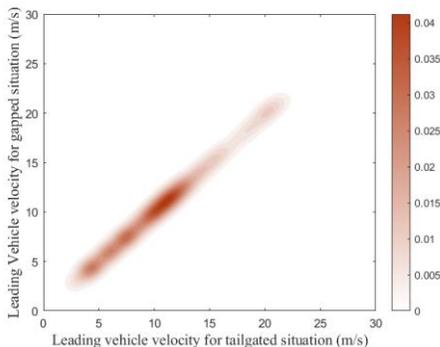

**Fig. 8.** Density heatmap for LV's speed in paired CF segments for different FV conditions.

## VI. Discussions

First, in line with previous CF models [4, 5, 9, 27], we found that drivers adjusted their CF behaviors based on the speed of the traffic flow and the relative distance to the LV. As shown in Figure 9 and aligning with the kernel density distribution in Figure 4, the range of positive rewards for spacing increased with the increase of the LV speed. In other words, at lower speeds, drivers tended to maintain shorter spacing, which ensures that the time headway does not become excessively large, thereby preserving driving efficiency.

Next, we have observed a strong influence of tailgated FV on the CF behaviors of the ego-vehicle. As observed in Figure 9, regardless of the speed of LV, the range of positive rewards distribution for spacing was consistently larger in the gapped CF events compared to the tailgated CF events. This result aligns with the findings from the kernel density distribution of paired driving segments in Figure 4, where the density distribution of spacing was broader when the FV was in a gapped state. At the same time, when the FV was in a tailgated state, positive rewards distribution were primarily observed when the relative speed was negative, indicating that the ego-vehicle was more inclined to approach the LV due to the pressure from the FV. In contrast, when the FV was in a gapped state, even when the LV speed was low (e.g., 4.3m/s), positive rewards were still distributed at positive relative speeds.

Additionally, it can be observed that, regardless of the LV speed and spacing between the ego-vehicle and LV, when the FV was in a tailgated state, the distribution range of positive rewards for the relative speed was comparably narrower compared to that in the gapped condition, indicating that when being tailgated, the ego-vehicle is more inclined to maintain a speed similar to that of LV. All the above-mentioned results suggest that when the FV exerted less pressure, the ego-vehicle had more freedom to choose its preferred driving style. Conversely, in a tailgated CF event, the driver might have experienced pressure from the FV, nudging them to maintain a shorter distance to the LV. These results are in line with the statistical analysis of the driving performance metrics. Specifically, the tailgating of the FV can lead to a closer CF distance of the ego-vehicle, though they would be more cautious at the same time.

Furthermore, we found that even when being tailgated, drivers still adjusted their CF strategies in response to the speed variation of the traffic flow. Specifically, when the LV speed was low (4.3 m/s and 7.4 m/s), the distribution range of positive rewards was concentrated in the negative part of the relative speed coordinate axis in the tailgated events. As the spacing increased, the distribution range of positive rewards increasingly moved to the negative side on the relative speed axis and eventually disappeared, reflecting the driver's tendency to get closer to the LV when the spacing became large. However, when the LV speed was high (11 m/s and 20 m/s), and when the spacing between the LV and the ego vehicle was small, drivers considered more than spacing; they also began to favor positive relative speed value to increase spacing. This indicates that as the speed of traffic flow increased, drivers assigned more weights on safety. Similar trend was also observed under the gapped condition, indicating that the perceived risk ahead may have shadowed the effects of the peer pressure from behind.

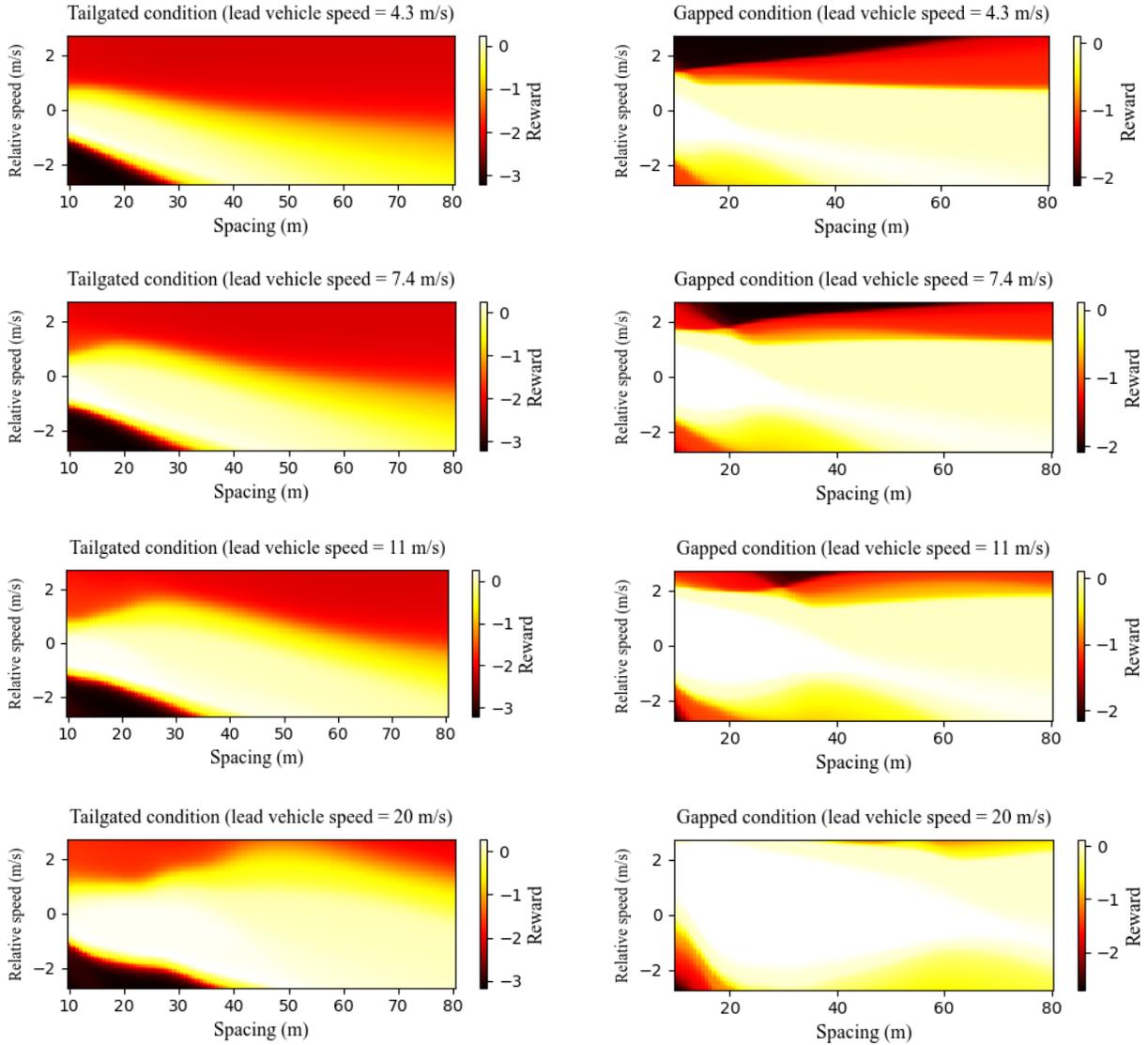

**Fig. 9.** Reward for ego-vehicle CF behavior at fixed LV speed in different FV conditions.

The above-mentioned findings suggest that drivers could adjust their preferred following strategy to some extent based on the speed of the traffic, though they still preferred a smaller spacing to the LV when being tailgated. These results may partially explain the insignificant results of the speed fluctuation metrics. In other words, drivers' speed choice is still mainly affected by the speed of the traffic flow, even though they can perceive the pressure from the FVs.

## VII. LIMITATIONS

Though, for the first time, we have observed real drivers' responses to tailgaters in an observational dataset, there are still some limitations. First, we only considered one dataset collected on highway. Future studies should explore more datasets in more diverse traffic scenarios. Second, we only considered driving performance data. It would be more interesting to delve into drivers' psychological states and cognitive processes when handling the tailgaters in CF events. Future research can integrate physiological data, eye-tracking and subjective responses to gain a better understanding of how the FV affects their own CF decisions.

## VIII. CONCLUSION

In summary, in our study, we investigated the influence of different FV conditions on the CF behavior of the ego-vehicle. Based on the time headway between the ego-vehicle and the FV, we categorized two FV conditions, i.e., tailgated and gapped. Then, we conducted analyses based on 1,024 tailgated CF events and 465 gapped CF events extracted from a naturalistic driving dataset, the highD dataset. We found that:

- In line with previous CF models, the states of the LV had an impact on the CF behaviors of the ego-vehicle.
- The tailgater had a strong influence on the ego-vehicle's CF strategies, leading to a shorter headway distance to the LV and less freedom in selecting CF behaviors.
- Though being tailgated, the drivers still adopted different CF strategies to follow the LV at different speeds and when the LV was at different distances, indicating that the drivers were still capable of balancing the peer pressure and safety to some extent.

In general, our study highlights the role of FV in shaping drivers' behavior and underscores the necessity of considering FV influences in CF models.

ACKNOWLEDGMENT

This work was supported by the Scientific Research Projects for the Higher-educational Institutions of Guangzhou (No. 2024312135), and in part by the Guangzhou Municipal Science and Technology Project (No. 2023A03J0011), and Guangdong Provincial Key Lab of Integrated Communication, Sensing and Computation for Ubiquitous Internet of Things (No.2023B1212010007).

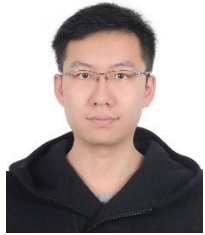

**Yang Liu** received the bachelor's degree in energy and power engineering from Harbin Engineering University, Harbin, China, in 2019, the M.S. degree in mechanical engineering from University of Alberta, Edmonton, Canda, in 2023. He is currently a Ph.D. student at the Thrust of Intelligent Transpiration, The Hong Kong University of Science and Technology (Guangzhou), Guangzhou, China.

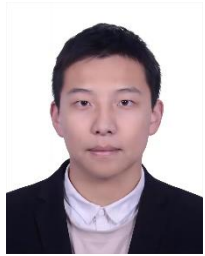

**Jiahao Zhang** received the B.S. degree from school of computer science and engineering, Sun Yat-sen University, Guangzhou, China, in 2023. He is currently a MPhil student at the Thrust of Intelligent Transpiration, The Hong Kong University of Science and Technology (Guangzhou), Guangzhou, China.

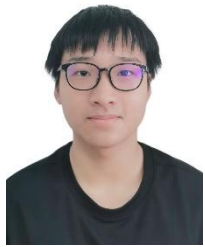

**Yuxuan Ouyang** is currently an undergraduate student at the college of educational science, The Hong Kong University of Science and Technology (Guangzhou), Guangzhou, China.

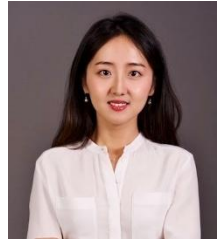

**Huan Yu** received the B.Sc. degree from Northwestern Polytechnical University, and the M.Sc. and Ph.D. degrees in mechanical and aerospace engineering from the University of California, San Diego, San Diego. She is an Assistant Professor with the Intelligent Transportation Thrust and the Robotics and Autonomous Systems Thrust, The Hong Kong University of Science and Technology (HKUST), Guangzhou, and an affiliate Assistant Professor with the Department of Civil and Environmental Engineering, HKUST. Her research has been focused on finding safe, stable and robust solutions that combine control theory, machine learning, and traffic flow theory to advance the boundaries of intelligent transportation systems.

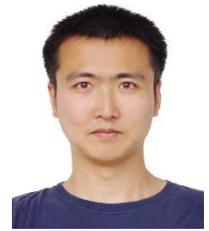

**Dengbo He** received the bachelor's degree in vehicle engineering from Hunan University, Changsha, China, in 2012, the M.S. degree in mechanical engineering from Shanghai Jiao Tong University, Shanghai, China, in 2016, and the Ph.D. degree from the University of Toronto, Toronto, ON, Canada, in 2020. He is currently an Assistant Professor with the Thrust of Intelligent Transpiration and Thrust of Robotics and Autonomous Systems, The Hong Kong University of Science and Technology, Guangzhou, China.